\documentclass[12pt]{cernart}
\usepackage{times}
\usepackage{graphicx}
\newcommand{\qx }{$q(x)$}
\newcommand{\Deqx }{$\Delta q(x)$}
\newcommand{\Deqtx }{$\Delta_T q(x)$}

\newcommand{\Deqt }{$\Delta_T q$}

\newcommand{\Cffun }{$\Delta_T^0 D_q^h$}
\newcommand{\gevc }{GeV/$c$}
\usepackage{rotating}
\begin{document}
\begin{titlepage}
\docnum{CERN--PH--EP/2005--003}
\date{10 February 2005}
%------------------------------------------------------

\vspace{1cm}
\title{\Large First Measurement of the Transverse Spin Asymmetries of the Deuteron
        in Semi-Inclusive Deep Inelastic Scattering}
%\title{\Large First Measurements of Collins and Sivers Deuteron Asymmetries}
%\title{\Large Measurement of the Transverse Asymmetry of the Deuteron}

\author{\large The COMPASS Collaboration}

\vspace{3cm}
\begin{abstract}
First measurements of the Collins and Sivers asymmetries of charged
hadrons produced in deep-inelastic scattering of muons
on a transversely  polarized $^6$LiD target are presented.
The data were taken in 2002 with the COMPASS spectrometer
using the muon beam of the CERN SPS at 160~\gevc .
The Collins asymmetry turns out to be
 compatible with zero, as does the measured Sivers asymmetry
within the present statistical errors.
\vfill
\submitted{(Submitted to Physical Review Letters)}
\end{abstract}

\newpage
\begin{Authlist}
{\large  The COMPASS Collaboration}\\[\baselineskip]
V.Yu.~Alexakhin\Iref{dubna},
Yu.~Alexandrov\Iref{moscowlpi},
G.D.~Alexeev\Iref{dubna},
A.~Amoroso\Iref{turin},
B.~Bade\l ek\Iref{warsaw},
F.~Balestra\Iref{turin},
J.~Ball\Iref{saclay},
G.~Baum\Iref{bielefeld},
Y.~Bedfer\Iref{saclay},
P.~Berglund\Iref{helsinki}
% ...
\end{Authlist}
%%%%%%%%%%%%%%%%%%%%%%%%%%%%%%%%%%%%%%%%%%%%%%%%%%%%%%%%%%%%%%%%%%%%%%%%%%%%%%%%%%%%%%%%%%%%%%%%%%%%%%%%%%%%%%%%%%%%%%%%%%%%%%%
%
% define institutes
%
%%%%%%%%%%%%%%%%%%%%%%%%%%%%%%%%%%%%%%%%%%%%%%%%%%%%%%%%%%%%%%%%%%%%%%%%%%%%%%%%%%%%%%%%%%%%%%%%%%%%%%%%%%%%%%%%%%%%%%%%%%%%%%%
\Instfoot{bielefeld}{Universit\"at Bielefeld, Fakult\"at f\"ur Physik, 33501 Bielefeld, Germany}
\Instfoot{bochum}{Universit\"at Bochum, Institut f\"ur Experimentalphysik, 44780 Bochum, Germany}
\Instfoot{bonniskp}{Universit\"at Bonn, Helmholtz-Institut f\"ur Strahlen- und Kernphysik, 53115 Bonn, Germany}
\Instfoot{bonnpi}{Universit\"at Bonn, Physikalisches Institut, 53115 Bonn, Germany}
\Instfoot{burdwan}{Burdwan University, Burdwan 713104, India}
\Instfoot{calcutta}{Matrivani Institute of Experimental Research \& Education, Calcutta-700 030, India}
\Instfoot{dubna}{Joint Institute for Nuclear Research, 141980 Dubna, Moscow region, Russia}
\Instfoot{erlangen}{Universit\"at Erlangen--N\"urnberg, Physikalisches Institut, 91058 Erlangen, Germany}
\Instfoot{freiburg}{Universit\"at Freiburg, Physikalisches Institut, 79104 Freiburg, Germany}
\Instfoot{cern}{CERN, 1211 Geneva 23, Switzerland}
\Instfoot{heidelberg}{Universit\"at Heidelberg, Physikalisches Institut,  69120 Heidelberg, Germany}
\Instfoot{helsinki}{Helsinki University of Technology, Low Temperature Laboratory, 02015 HUT, Finland 
                             and University of Helsinki, Helsinki Institute of Physics, 00014 Helsinki, Finland}
\Instfoot{mainz}{Universit\"at Mainz, Institut f\"ur Kernphysik, 55099 Mainz, Germany}
\Instfoot{moscowlpi}{Lebedev Physical Institute, 119991 Moscow, Russia}
\Instfoot{munichtu}{Technische Universit\"at M\"unchen, Physik Department, 85748 Garching, Germany}
\Instfoot{munichlmu}{Ludwig-Maximilians-Universit\"at M\"unchen, Department f\"ur Physik, 80799 Munich, Germany}
\Instfoot{nagoya}{Nagoya University, 464 Nagoya, Japan}
\Instfoot{protvino}{State Research Center of the Russian Federation, Institute for High Energy Physics, 142281 Protvino, Russia}
\Instfoot{saclay}{CEA DAPNIA/SPhN Saclay, 91191 Gif-sur-Yvette, France}
\Instfoot{telaviv}{Tel Aviv University, School of Physics and Astronomy, 69978 Tel Aviv, Israel}
\Instfoot{triestictp}{ICTP--INFN MLab Laboratory, 34014 Trieste, Italy}
\Instfoot{triest}{INFN Trieste and University of Trieste, Department of Physics, 34127 Trieste, Italy}
\Instfoot{turin}{INFN Turin and University of Turin, Physics Department, 10125 Turin, Italy}
\Instfoot{warsaw}{So\l tan Institute for Nuclear Studies and Warsaw University, 00-681 Warsaw, Poland }
\Instfoot{warsawtu}{Warsaw University of Technology, Institute of Radioelectronics, 00-665 Warsaw, Poland }
\Anotfoot{a}{Also at University of East Piedmont, 15100 Alessandria, Italy}
\Anotfoot{b}{On leave of absence from JINR Dubna}
\vfill

\hbox to 0pt {~}
\end{titlepage}
%
%
%\begin{keyword}
%deep inelastic scattering \sep structure functions
%\PACS{13.60.Hb \sep 13.88.+e}
%\pacs{13.60.-r, 13.88.+e, 14.20.Dh, 14.65.-q}
%\end{keyword}
%
%=================================================================================================================

The importance of transverse spin effects at high energy
in hadronic physics was first suggested by the discovery in 1976 that
$\Lambda$ hyperons produced in $pN$ interactions exhibited 
an anomalously large transverse polarization~\cite{Bunc76}. 
This effect could not be easily explained.
For a long time it was believed to be forbidden 
at leading twist in QCD~\cite{Kane78}, and
very little theoretical work was
devoted to this field for more than a decade.

This situation changed in the nineties. 
After the first hints of large single transverse spin asymmetries 
in inclusive $\pi^0$ production in polarized pp scattering at CERN~\cite{cern},
remarkably large asymmetries were found at 
Fermilab both for neutral and charged pions~\cite{E704a}. 
In parallel, intense theoretical activity was taking
place: the significance of the quark transversity distribution,
already introduced in 1979~\cite{RaSo79} to describe a quark in a
transversely polarized nucleon, was reappraised~\cite{ArMe90} in 1990, 
and its measurability via the Drell--Yan
process established. 
In 1991 a general scheme of all leading
twist and higher-twist parton distribution functions 
was worked out~\cite{JaJi91}, and in
1993 a way to measure
transversity in lepton nucleon polarized deep-inelastic scattering (DIS)
was suggested~\cite{Coll93}.
On the experimental
side, the RHIC-Spin Collaboration~\cite{RHIC} and
the HELP Collaboration~\cite{HELP} put forward the first proposals to 
measure transversity.
Today transversity is an
important part of the scientific programme of the HERMES experiment
at DESY and of the COMPASS experiment at CERN, both presently taking data.
First results on a transversely polarized proton target have been published
recently by the HERMES Collaboration~\cite{Hermest}.

To fully specify  the quark structure of the nucleon 
at the twist-two level, 
the transverse spin distributions \Deqtx\  must be added to the momentum
distributions \qx\ and the helicity distributions \Deqx ~\cite{JaJi91}.
For a discussion on notation, see Ref.~\cite{BDR02}.
If the quarks are collinear with the parent nucleon (no intrinsic quark
transverse momentum $k_T$), or after integration over $k_T$,
these three distributions exhaust the information
on the internal dynamics of the nucleon.
More distributions are allowed admitting a finite  % \kt\ 
$k_T$, or at higher twist~\cite{JaJi92,aram94,MuTa96,BDR02}.

\begin{figure}[t] % 
\centerline{\includegraphics[width=8.2cm]{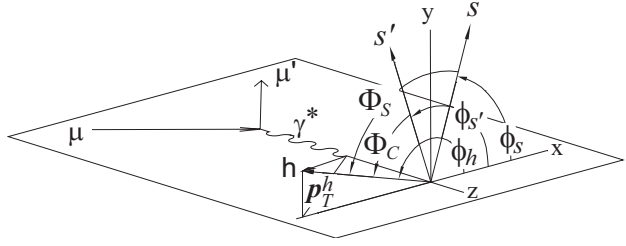}}
\caption{Definition of the Collins and Sivers angles.}
\label{fig:colla}
\end{figure}

The distributions \Deqt\ are difficult to measure, since 
they are chirally odd and therefore absent in
inclusive DIS. They  may instead be
extracted from measurements of the single-spin asymmetries in cross-sections 
for semi-inclusive DIS (SIDIS)
of leptons on transversely polarized nucleons, in which 
a hadron is also detected in the final state.
In these processes the measurable asymmetry,
the ``Collins asymmetry'' $A_{Coll}$, is due to the combined effect 
of \Deqt\ and another chirally-odd function, \Cffun , which
describes the spin-dependent part of the hadronization of a transversely 
polarized quark $q$ into a hadron $h$. At leading order 
in the collinear case $A_{Coll}$ can be written as 
\begin{eqnarray}
A_{Coll} = \frac {\sum_q e_q^2 \cdot \Delta_T q \cdot \Delta_T^0 D_q^h}
{\sum_q e_q^2 \cdot q \cdot D_q^h}
\label{eq:collass}
\end{eqnarray}
where $e_q$ is the quark charge.
According to Collins~\cite{Coll93}, the quantity \Cffun\ can be obtained by
investigating the fragmentation of a polarized quark
$q$ into a hadron $h$, and is related to the $\vec{p}_T^{\, h}$ 
dependent fragmentation function
\begin{eqnarray}
D_{T\,q}^{\; \; \; h}(z, \vec{p}_T^{\, h}) = D_q^h(z, | \vec{p}_T^{\, h}|^2) + 
    \Delta_T^0 D_q^h(z, | \vec{p}_T^{\, h}|^2) \cdot \sin\Phi_{C} .
\label{eq:collfun}
\end{eqnarray}
Here
$\vec{p}_T^{\, h}$ is the hadron transverse momentum
with respect to the struck quark direction, i.\,e.~the virtual 
photon direction, and
%\linebreak
$z = E_h / (E_{l}-E_{l'})$ is the fraction of available energy 
carried by the hadron.
$E_h$, $E_{l}$, and $E_{l'}$ are the energies of the hadron, the incoming lepton,
 and the scattered lepton respectively.
The ``Collins angle'' $\Phi_{C}$ is conveniently defined in a coordinate
system in which the z-axis is the  virtual photon direction
and the x-z plane is the lepton scattering plane, as illustrated in 
Fig.~\ref{fig:colla}. 
In this reference system $\Phi_{C}= \phi_h -\phi_{s'}$, where
$\phi_h$ is the azimuthal angle of the hadron, and $\phi_{s'}$ is the azimuthal angle
of the transverse spin of the struck quark. 
Since $\phi_{s'}=\pi-\phi_s$, with $\phi_s$ the azimuthal 
angle of the transverse 
spin of the initial quark (nucleon), one obtains
$\sin \Phi_{C} = - \sin(\phi_h +\phi_s)$.

An entirely different mechanism was suggested by Sivers~\cite{Sivers} 
as a possible cause of the transverse spin effects observed in pp
scattering. 
This mechanism could also be responsible for a spin asymmetry in the cross-section of SIDIS 
of leptons on transversely polarized nucleons.
Allowing for an intrinsic $\vec{k}_T$ dependence of the quark distribution
in a nucleon, a left-right asymmetry could 
be induced in such a distribution by a transverse nucleon polarization, 
%%%%%\linebreak
%%%%%\pagebreak
$q_{T}(x,\vec{k}_T)= q(x,|\vec{k}_T|^2) + \Delta_0^T q(x,|\vec{k}_T|^2)
\cdot \sin \, \Phi_{ S}$, where $\Phi_{ S}= \phi_h -\phi_s \neq \Phi_{C}$ is the
``Sivers angle''.
Neglecting the hadron transverse momentum with respect to the
fragmenting quark, this $\vec{k}_T$ dependence could cause the ``Sivers
asymmetry'' 
\begin{eqnarray}
A_{Siv} = \frac {\sum_q e_q^2 \cdot \Delta_0^T q \cdot D^h_q}
{\sum_q e_q^2 \cdot q \cdot D_q^h} 
\label{eq:sivass}
\end{eqnarray}
in the distribution of the hadrons resulting from the quark
fragmentation with respect to the nucleon polarization which
could be revealed as a $\sin \Phi_{ S}$ modulation
in the number of produced hadrons.
Measuring SIDIS on a transversely polarized target allows the 
Collins and the Sivers effects to be disentangled~\cite{colsiv}.

In this paper first results are given of the charged hadron single-spin
asymmetries in SIDIS of high energy muons on a transversely polarized $^6$LiD 
target measured in 2002 by the COMPASS Collaboration.

The COMPASS spectrometer has been set up at the CERN SPS muon beam. 
The experiment has taken data from 2002 to 2004 
%The experiment started taking data in 2002 
%and up to 2004 has been run 
at a muon momentum of 160~\gevc\
with beam rates of $4 \cdot 10^7$ muons/s.
The beam is naturally polarized by the $\pi$-decay mechanism,
with a polarization of about $-76$\%.
The polarized target system~\cite{Ctarget}
consists of two  cells (upstream $u$, downstream $d$), each 60 cm long,
located along the beam one after the other in two separate RF cavities,
and oppositely polarized.
The target magnet can provide both a solenoid field (2.5~T), and a dipole field
(0.4~T) used for adiabatic spin rotation and for the transversity measurements.
Correspondingly, the target polarization can be oriented either longitudinally
or transversely to the beam direction.
Polarizations of 50\% have been reached routinely with the $^6$LiD target,
which has a favorable dilution factor $f \simeq 0.4$, since $^6$Li
basically consists of a deuteron plus an $^4$He core.
The target polarization is measured with a relative precision of 5\%.
Particle tracking is performed using several
stations of scintillating fibers, micromesh gaseous
chambers, and gas electron multiplier chambers.
Large-area tracking devices comprise gaseous detectors
(drift chambers, straw tubes, and MWPCs) placed around the two spectrometer 
magnets. 
Muons are identified in large-area Iarocci tubes and drift 
tubes downstream of hadron absorbers.
The trigger~\cite{trigger} is formed by several hodoscope 
systems supplemented by two hadron 
calorimeters. Veto counters are installed in front 
of the target to reject the beam halo.
More information on the COMPASS spectrometer can be found in Ref.~\cite{FB03}.

In 2002 about $6 \cdot 10^9$ events, corresponding to 
260~TBytes of data, were collected. 
About 20\% of the sample was taken 
in the transverse spin mode, in two separate periods.
Each period started with the \textit{u}-cell of the target downwardly polarized 
and the \textit{d}-cell upwardly polarized. 
After 4--5 days a  polarization reversal was performed 
by changing the RF frequencies in the two cells.

Because the asymmetries are obtained by comparing data taken 
several days apart, the stability of the apparatus is crucial.
To check the stability of reconstruction, the data were sampled in time. 
The hit
distributions on all trackers were scrutinized, as well as the number of 
reconstructed events, the number of
vertices per event,  and the number of tracks per
event in the whole spectrometer and in its various subregions.
In addition, the distributions of a few relevant quantities were
monitored for their stability throughout the data, like the Bjorken 
variable $x$, the relative energy transfer in the muon scattering
process $y = (E_{l}-E_{l'})/E_{l}$, the photon virtuality $Q^2$.
%the transverse momentum $p_T^h$, the laboratory azimuthal angle
%$\phi_h^{lab}$ of the most energetic hadron in the event, and the
%accuracy of the reconstructed $\rm{K^0_s}$ mass. 
These investigations led to the exclusion of
%%%%some 4\% of the  data from the final data set.
about 4\% of the  data from the final sample.

In the analysis, events were selected in  which a vertex with incident and
scattered muon and at least one outgoing charged hadron was found in one 
of the two target cells. 
A clean identification of muons and hadrons was achieved 
on the basis of the amount of material traversed in the spectrometer.
In addition, DIS cuts $Q^2 > 1$~(\gevc )$^2$, 
$W > 5$~\gevc $^2$, and $0.1 < y < 0.9$ were
applied to the data as well as a cut on the transverse momentum
of the hadrons ($p_T^h > 0.1$ \gevc ).

To enhance the asymmetry signal, we first evaluated 
the Collins and Sivers asymmetries for the 
leading hadron of each event, the underlying idea being that in the string 
fragmentation it is the most sensitive to the properties of the parent 
quark spin~\cite{Artru}.
The leading hadron was defined as the most energetic hadron with $z > 0.25$,
and originating from the reaction vertex. 
The total number of events which finally entered the analysis 
was $1.6 \cdot 10^6$ comprising $8.7 \cdot 10^5$  events with positive 
leading hadrons and $ 7.0 \cdot 10^5$ events with negative leading hadrons.

We searched separately for Collins and Sivers asymmetries in the data.
The $\Phi$ distribution
of the number of events for each cell and for each 
polarization state can be written as
\begin{eqnarray} \hspace*{-0.6cm}
N_{j}(\Phi_j) & = & F \, n  \, \sigma \cdot a_{j}(\Phi_j) 
\cdot (1 + \epsilon_{j} \, \sin \Phi _j) ,
%N_{h,{C(S)}(\Phi_{C(S)}) &\propto& a_{C(S)}(\Phi_{C(S)}) 
%\cdot (1 + \epsilon_{C(S)} \cdot \sin \Phi_{C(S)}); \; j=C,\, S \;\;\,
\label{eq:angdis}
\end{eqnarray}
where $j=C, S$, and $F$ is the muon flux, $n$ the number of target particles,
$\sigma$ the spin averaged cross-section, and $a_{j}$ the product of angular acceptance 
and efficiency of the spectrometer.
The asymmetries $\epsilon_j$ are
%\linebreak
$\epsilon_C = f \cdot |P_T| \cdot D_{NN} \cdot A_{Coll}$ and 
$\epsilon_S = f \cdot |P_T| \cdot A_{Siv}$. 
%\linebreak
The factor $f$ is the polarized target dilution factor, 
$P_T$ the deuteron polarization, and
$D_{NN} = (1-y)/(1-y+y^2/2)$ the transverse spin
%\linebreak
transfer coefficient from the initial to 
the struck quark~\cite{BDR02}.
To highlight the physics process we are after, in
Eq.~\ref{eq:angdis} we have omitted terms which either average out
in the evaluation of the asymmetry or only lead to negligible
corrections due to a non-uniform angular acceptance.
The beam polarization contributes to the  asymmetry
only by higher-twist effects, which are not considered in this leading-order 
analysis.

The asymmetries $\epsilon_C$ and $\epsilon_S$
were evaluated from the number of events with
the two target spin orientations
($\uparrow$ spin up, and $\downarrow$ spin down)
by fitting the quantities
\begin{eqnarray}
A_{j}^m(\Phi_{j}) = \frac{N^{\uparrow}_j(\Phi_{j}) 
                      - r \cdot N^{\downarrow}_j(\Phi_{j}+ \pi)}
{N^{\uparrow}_j(\Phi_{j})
                       + r \cdot N^{\downarrow}_j(\Phi_{j}+ \pi)}
\label{eq:asim}
\end{eqnarray}
with the functions
$\epsilon_C \cdot \sin \Phi_{C}$ and 
$\epsilon_S \cdot \sin \Phi_{S}$.
The normalization factor $r$ has been taken equal to the ratio
%\linebreak
%$r = N^{\uparrow}_{h, tot}/N^{\downarrow}_{h, tot}$ is the ratio 
of the total number
of detected events in the two orientations of the target polarization.
Note that two events having the same topology in the laboratory
before and after the target spin rotation have 
angles $\Phi_{j}$ and $\Phi_{j}+ \pi$ respectively, thus the
acceptance cancels in Eq.~\ref{eq:asim} as long as the ratio
$a_j^{\uparrow}(\Phi_{j}) / a_j^{\downarrow}(\Phi_{j}+ \pi)$
is constant in $\Phi_{j}$.

\begin{figure*}[t] % 
%\begin{minipage}[w]{18cm}
\centerline{\includegraphics[width=\hsize]{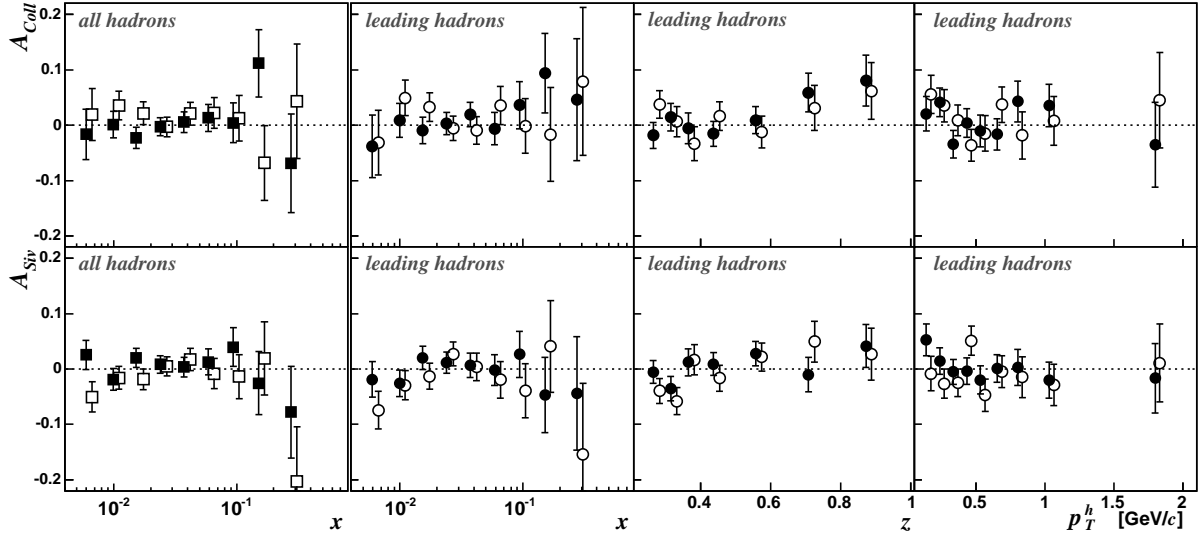}}
\caption{\label{results} Collins asymmetry (top) and Sivers asymmetry 
(bottom) against $x$, $z$ and 
$p_T^h$ for positive (full points) and negative hadrons (open points).
Error bars are statistical only.
The first column gives the asymmetries for all hadrons,
the other three columns for the leading hadrons.
In all the plots the points are slightly shifted 
horizontally with respect to the measured value.}
%\end{minipage}
\end{figure*}

The evaluation of the asymmetries was performed separately for the 
two data-taking periods and for the two target cells. 
These four sets of measured asymmetries turned out to be statistically
compatible, and were then combined by taking weighted averages.
Plots of the measured values of 
%Collins and Sivers asymmetries 
$A_{Coll}$ and $A_{Siv}$ against the three kinematic
variables $x$, $z$ and $p_T^h$ are given in Fig.~\ref{results}. 
The errors shown in the figure are only statistical. 
The mean values of $z$ and $p_T^h$ are roughly constant
($\sim 0.44$  and 0.51~\gevc\ respectively) over the whole $x$ range
while $\langle Q^2 \rangle$ increases 
from $\sim 1.1$ (\gevc )$^2$ in the first $x$ bin to
$\sim 20$ (\gevc )$^2$ in the last one.

Systematic errors due to the uncertainties in $P_T$, $D_{NN}$,
and $f$ are negligibly small.
Several tests were made to check that there are no effects distorting 
the measured asymmetries, splitting the data sample 
$i)$ in time,
$ii)$ in  two halves of the target cells, and
$iii)$ according to the hadron momentum. 
The asymmetries measured for the different samples were found
to be compatible.
Also, the results were stable with respect to different choices of
the normalization factor $r$.

The method of extracting the asymmetries is expected to
minimize systematic effects due to acceptance,
and this is confirmed by the compatibility of the asymmetries
measured in the two cells $u$ and $d$.
Under the reasonable assumption that the ratio
$a_{j,u}^{\downarrow}(\Phi_{j}+ \pi) / a_{j,d}^{\uparrow}(\Phi_{j})$ 
before the polarization reversal be equal to the corresponding ratio
$a_{j,u}^{\uparrow}(\Phi_{j}) / a_{j,d}^{\downarrow}(\Phi_{j}+ \pi)$
after the reversal, the requirement that the ratios 
$a_{j,u}^{\downarrow}(\Phi_{j}+ \pi) / a_{j,u}^{\uparrow}(\Phi_{j})$ and
$a_{j,d}^{\uparrow}(\Phi_{j}) / a_{j,d}^{\downarrow}(\Phi_{j}+ \pi)$,
be constant in $\Phi_{j}$ within each data-taking 
period has been verified by constructing the ratio 
\begin{eqnarray}
R_j(\Phi) = \frac{ N_{j,u}^{\uparrow}(\Phi_{j}) 
                      \cdot N_{j,d}^{\downarrow}(\Phi_{j}+ \pi)} 
                     {N_{j,u}^{\downarrow}(\Phi_{j}+ \pi) 
                               \cdot N_{j,d}^{\uparrow}(\Phi_{j})}
\propto \frac{ [
a_{j,u}^{\uparrow}(\Phi_{j}) ]^2 }{ [ a_{j,u}^{\downarrow}(\Phi_{j}+ \pi)]^2}
\label{eq:ratior}
\end{eqnarray}
and verifying its constancy in $\Phi_{j}$.
This constancy holds even using the entire data sample after releasing the $z$-cut.
It has to be stressed also that, under the same assumption,
possible false asymmetries due to variations in $\Phi_{j}$
of the acceptance ratios to first order have opposite sign in the two cells 
and  cancel in the average.

To estimate the size of possible systematic effects, 
the asymmetries have also been evaluated using two other estimators
which are
independent of relative luminosities and rely on different assumptions
of the acceptance variations, e.g.~the ratio product
\begin{eqnarray}
\frac{ N_{j,u}^{\uparrow}(\Phi_{j})}
                     {N_{j,u}^{\downarrow}(\Phi_{j}+ \pi)} \cdot
\frac{ N_{j,d}^{\uparrow}(\Phi_{j})} 
                     {N_{j,d}^{\downarrow}(\Phi_{j}+ \pi)} ,
\label{eq:estf}
\end{eqnarray}
and the geometric mean
\begin{eqnarray}
   \frac{\sqrt{ N_{j}^{\uparrow}(\Phi_{j})
     \cdot N_{j}^{\downarrow}(\Phi_{j})}
                           -\sqrt{ N_{j}^{\downarrow}(\Phi_{j}+\pi)
     \cdot N_{j}^{\uparrow}(\Phi_{j}+\pi)}}
                           {\sqrt{ N_{j}^{\uparrow}(\Phi_{j})
     \cdot N_{j}^{\downarrow}(\Phi_{j})}
                           +\sqrt{ N_{j}^{\downarrow}(\Phi_{j}+\pi)
     \cdot N_{j}^{\uparrow}(\Phi_{j}+\pi)}} .
\label{eq:estg}
\end{eqnarray}
Differences from the results displayed in Fig.~\ref{results}
were only observed within the 
statistical errors of the measured asymmetries.

The conclusion from all these studies is that systematic errors 
are smaller than the quoted statistical errors.

Within the statistical accuracy of the data, both 
%the Collins and the Sivers asymmetries on
$A_{Coll}$ and $A_{Siv}$ turn out to be small and compatible with zero, 
with a marginal indication of a Collins effect
at large $z$ in both the positive and the negative hadron data.
By means of Monte Carlo simulations, we estimated that 
the following factors could together dilute a possible leading pion asymmetry
by a factor of 0.6 at most: 
{\em  i)} the acceptance of the spectrometer for 
leading  hadrons (by cutting at $z > 0.25$ the reconstructed charged leading 
particle is the generated most energetic hadron in about 80\% of the cases);
{\em ii)} non identification of the charged hadron (about 80\% of the charged 
leading hadrons are pions); 
{\em iii)} smearing of the kinematical quantities due to the
experimental resolution of the spectrometer (negligible effect). 
For the simulation, which reproduces well the experimental distributions, 
we used LEPTO~6.5.1 and GEANT~3.
Simulations were also performed to check the possible correlation
between the measured values of $\epsilon_C$ and $\epsilon_S$;
asymmetries up to 20\% were generated and no appreciable mixing was
observed.

This analysis has been repeated for all hadrons, i.\,e.~both the
Collins and the Sivers asymmetries have been evaluated for all the 
reconstructed hadrons with $z > 0.2$.
The total number of hadrons entering the analysis is increased 
by a factor of 1.5 with respect to the leading hadron analysis, but the 
results are very similar, i.\,e.~small values for the asymmetries.
For reasons of space, the asymmetries  are displayed
in Fig.~\ref{results} as function of $x$ only.
All the measured asymmetries are available on HEPDATA~\cite{hepdata}.

The COMPASS measurements on the transversely polarized deuteron target
have a statistical accuracy of the same order as the recent 
measurement on protons performed by the HERMES 
Collaboration~\cite{Hermest}.
The small measured values of the deuteron asymmetries can be understood 
because $\Delta_T u$ and $\Delta_T d$ are likely to have the opposite sign
as for the helicity distributions, and some cancellation is expected between the 
proton and the neutron asymmetries.
Still, at large $x$, the measured values of $A_{Coll}$ for positive leading hadrons
seem to hint at positive values, at variance with the naive expectation
$A_{Coll}^{\pi^+} \propto - \Delta_T u / u$.
Also, $A_{Coll}$ for all positive hadrons does not show the negative trend
foreseen by the model prediction of Ref.~\cite{efremov}.
Attention is drawn to the fact that the conventions used in Ref.~\cite{Hermest}
and ~\cite{efremov} give an opposite sign for the Collins 
asymmetry as compared to this paper.
Alternatively, it could  be 
%that the Sivers mechanism is not effective, and 
that the Collins effect is too small to allow for quark polarimetry with this
set of data.
Different quark polarimeters are also being tried, e.\,g.~hadron pairs
and $\Lambda$ production.
The analysis of the full sample of deuteron data, including the 2003 and 2004
runs, will reduce the errors by at least a factor of two,
and the Collaboration also intends to take data with a polarized proton target.
Precise transversely polarized
proton and deuteron data will allow a flavor separation of transversity
in the near future.
%\\

\section*{Acknowledgements}
We acknowledge the support of the CERN management and staff, as well as
the skills and efforts of the technicians of the collaborating
institutes. This work is supported by MEYS (Czech Rep.), BMBF (Germany),
UGC-DSA and SBET (India), ISF (Israel), INFN and MIUR (Italy), MECSST (Japan),
KBN (Poland), FCT (Portugal).

%\bibliography{article}
%\bibliographystyle{unsrt}

%%%%%%%%%%%%%%%%%%%%%%%%%%%%%%%%%%%%%%%%%%%%%%%%%%%%%%%%%%%%%%%%

%
%%%%%%%%%%%%%%%%%%%%%%%%%%%%%%%%%%%%%%%%%%%%%%%%%%%%%%%%%%%%%%%%

\end{document}